\documentclass{JHEP3}

  \usepackage{latexsym,bm,amsmath,amssymb,amsfonts}
  \usepackage{epsfig,graphics,graphicx}
  \usepackage{slashed}
  \usepackage{cite}
  \usepackage[latin1]{inputenc}

  \long\def\comment#1{ }

  \newcommand{\beq}{\begin{eqnarray}}
  \newcommand{\eeq}{\end{eqnarray}}
  
 \def\simge{\mathrel{%
   \rlap{\raise 0.511ex \hbox{$>$}}{\lower 0.511ex \hbox{$\sim$}}}}
\def\simle{\mathrel{
   \rlap{\raise 0.511ex \hbox{$<$}}{\lower 0.511ex \hbox{$\sim$}}}}

\vspace{1.5cm}

\keywords{Hadronic Colliders, AdS-CFT Correspondence}
\preprint{
}

\title{\rm \LARGE Relating $e^+e^-$ annihilation to high energy scattering at weak and strong coupling}

\author{Yoshitaka Hatta\\Graduate School of Pure and Applied Sciences, University
of Tsukuba, Tsukuba, Ibaraki 305-8571, Japan\\
E-mail: \email
{hatta@het.ph.tsukuba.ac.jp
 }}

\abstract{
We explore the correspondence between the final state in $e^+e^-$ annihilation and the small--$x$ hadronic wavefunction in the transverse plane  both in weakly coupled QCD and strongly coupled ${\mathcal N}=4$ SYM. At strong coupling, the virtual and static photon produced in $e^+e^-$ annihilation can be treated as a shock wave propagating in AdS space leaving spherical energy and charge distributions on the boundary. This is shown to be mathematically identical to the computation of energy and charge distributions in the transverse plane generated by a high energy color singlet state.  At weak coupling, the correspondence is useful in studying interjet observables. By performing the stereographic projection to the BFKL equation, we construct an exact solution to the evolution equation derived by Marchesini and Mueller, and find the angular distribution of small--$x$ gluons in the interjet region. Finally we argue that the correspondence holds also for the energy correlation functions.
}

\begin{document}

\section{Introduction}
In recent years evidence has been accumulating that the BFKL dynamics \cite{Balitsky:1978ic} is relevant to certain observables in $e^+e^-$ annihilation \cite{Dasgupta:2002bw,Banfi:2002hw,Marchesini:2003nh,Weigert:2003mm,Marchesini:2004ne,Forshaw:2005sx}. This is a rather unexpected development since, as is well--known \cite{Bassetto:1984ik}, the dominant contribution to the total multiplicity in $e^+e^-$ annihilation comes from narrow jets whose structure is described by the double logarithmic (soft and collinear) resummation, whereas the BFKL Pomeron has solely to do with the soft radiation. However, if one turns to regions away from jets, gluons are not sensitive to the collinear singularity as they are necessarily emitted at large angle. The  resummation of the remaining soft logarithms then leads to evolution equations very similar to the ones derived in the context of BFKL and saturation physics. In particular, the interjet soft gluon multiplicity to leading order in $N_c$ grows exactly like the BFKL Pomeron, suggesting the possible existence of a deep correspondence between the timelike and spacelike processes.   Still, conceptual as well as technical differences persist, and it remains unclear whether this similarity could be elevated to something more fundamental.

Meanwhile, a surge of interest in the Pomeron \cite{Polchinski:2001tt,Kotikov:2004er,Brower:2006ea,
Cornalba:2006xk,Stasto:2007uv,Brower:2007qh,Cornalba:2007fs,
BallonBayona:2008zi,AlvarezGaume:2008qs,Hatta:2007he,Bartels:2008zy,Albacete:2008ze} and $e^+e^-$ annihilation \cite{Hofman:2008ar,Hatta:2008tx,Hatta:2008qx,hatta} in strongly coupled ${\mathcal N}=4$ supersymmetric Yang--Mills (SYM) theory based on the AdS/CFT duality \cite{Maldacena:1997re,Gubser:1998bc} has provided new perspectives on the corresponding QCD problems. We now know that jets exist only in weakly coupled theories. This in turn suggests that the importance of the soft, non--collinear radiation in $e^+e^-$ annihilation becomes more pronounced as the coupling gets stronger. In the limit of infinite 't Hooft coupling  $\lambda \to \infty$, there are no jets, and therefore in a sense the entire solid angle would look like an interjet region. In such a situation one would naively expect that the \emph{total} multiplicity could be dictated by the Pomeron. Interestingly, the existing results already hint at this possibility. The AdS/CFT prediction of the multiplicity in $e^+e^-$ annihilation is \cite{Hatta:2008tx,Hatta:2008qx}
\begin{align} n(Q) \propto Q^{1-3/2\sqrt{\lambda}}\,, \end{align} where $Q$ is the virtual photon energy. On the other hand, the Pomeron intercept is \cite{Kotikov:2004er,Brower:2006ea} \begin{align} J_{\scriptsize{\mbox{Pom}}}-1=1-\frac{2}{\sqrt{\lambda}}\,. \end{align} One sees that in the strong coupling limit\footnote{It is conceivable that the remnant of jets at finite 't Hooft coupling accounts for the small discrepancy of order  $1/\sqrt{\lambda}$ in the exponent between the total and `interjet' multiplicities.} it may be  possible to interpret the linear dependence $n(Q)\propto Q$ as arising from the Pomeron (or rather, the graviton) $n(Q)\propto  Q^{J_{\scriptsize{\mbox{Pom}}}-1}$.

In this paper we explore further connections of this kind, between the final state in $e^+e^-$ annihilation and the small--$x$ wavefunction of a color singlet state (`hadron') both at weak (BFKL Pomeron) and strong (graviton) coupling.   At strong coupling, gravity in the $\mbox{AdS}_5$ space determines the energy distribution of a high energy hadron in the transverse plane, i.e., the plane perpendicular to the direction of motion \cite{Gubser:2008pc}. On the other hand, the angular distribution of energy in $e^+e^-$ annihilation is spherical \cite{Hofman:2008ar}. We  show in Section~2 that these results are just two sides of the same coin--mathematically identical, but worked out in  different coordinate systems.
Section~3 deals with the weak coupling case. The BFKL equation, once reformulated in the dipole language \cite{Mueller:1993rr}, predicts not only the number of small--$x$ gluons, but also their distribution and correlation in the transverse plane. It turns out that they can be exactly mapped onto the angular distribution  of soft gluons in the interjet region in $e^+e^-$ annihilation. The key element that realizes this mapping--the stereographic projection--has been inspired by the coordinate transformation in $\mbox{AdS}_5$ just mentioned above.
 Finally, in Section~4 we argue that the correspondence holds also for the energy--energy correlation functions.

\section{Energy and charge distribution at strong coupling}
\setcounter{equation}{0}
In this section we compute the energy and charge distributions on a sphere for $e^+e^-$ annihilation and in the transverse plane for a high energy hadron at strong coupling in such a way that the symmetry between the two computations are manifest.

\subsection{Metric conventions}
The $\mbox{AdS}_5$ space in the global coordinates is a hypersurface parameterized by the equation
\begin{align} W_{-1}^2+W_0^2-W_1^2-W_2^2-W_3^2-W_4^2=R^2\,. \end{align}
We set the AdS radius $R=1$ henceforth and introduce two Poincar\'e coordinate systems:\\

\underline{Poincar\'e 1}
\begin{align} W_{-1}+W_4=\frac{1}{z}\,, \quad W_\mu=\frac{x^\mu}{z}\,. \quad (\mu=0,1,2,3) \end{align}  The proper length is given by
\begin{align} ds^2=\frac{dz^2-2dx^+dx^- +d\vec{x}_T^2}{z^2}\,. \end{align}
 where $x^\pm=(x^0\pm x^3)/\sqrt{2}$ and $\vec{x}_T=(x^1,x^2)$. This is the natural coordinate system to describe high energy reactions involving  hadrons moving in the $x^3$--direction in the initial state. \\

\underline{Poincar\'e 2}
\begin{align} W_0+W_3=\frac{1}{y_5}=&\frac{-1}{\sqrt{2}y^+z}\,,\quad W_{-1}=-\frac{y^0}{y_5}\,,\quad W_4=-\frac{y^3}{y_5}\,, \quad W_{1,2}=\frac{y^{1,2}}{y_5}\,, \\
&ds^2=\frac{dy_5^2-2dy^+dy^- +d\vec{y}_T^2}{y_5^2}\,, \end{align} with $y^\pm=(y^0\pm y^3)/\sqrt{2}$ and $\vec{y}_T=(y^1,y^2)$. This coordinate system was previously introduced in \cite{Cornalba:2007fs} in the context of high energy scattering. As demonstrated in \cite{Hofman:2008ar}, it is particularly convenient for the discussion of the final state in $e^+e^-$ annihilation.\footnote{Our convention differs from \cite{Hofman:2008ar} by factors of $\sqrt{2}$.}

The following relations hold  on the boundary of $AdS_5$, at $y_5=0$ or $z=0$ \begin{align} y^+=-\frac{1}{2x^+}, \quad y^-=x^- -\frac{x_1^2+x_2^2}{2x^+}, \quad \vec{y}_T=\frac{\vec{x}_T}{\sqrt{2}x^+}\,. \end{align}
 The four--dimensional proper length is \begin{align}
ds_4^2= -2dx^+dx^-+ d\vec{x}_T^2=\frac{-2dy^+dy^- + d\vec{y}_T^2}{2(y^+)^2}\,. \label{bou} \end{align}

\subsection{A shock wave picture of $e^+e^-$ annihilation}

\begin{figure}
\begin{center}
\centerline{\epsfig{file=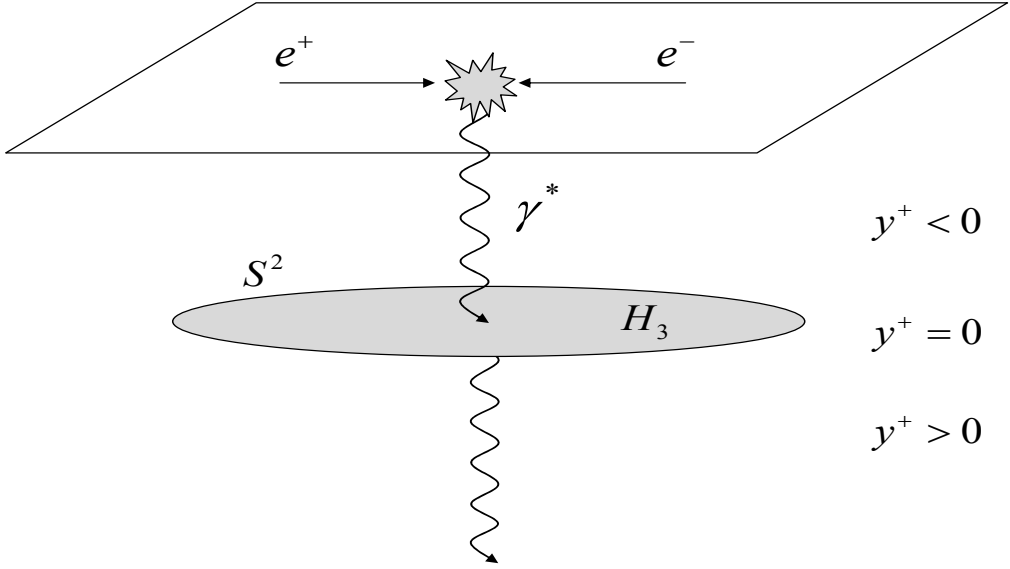,height=7.5cm,width=13.cm}}
\caption{\sl
 A schematic picture of $e^+e^-$ annihilation at strong coupling.
 \label{fig0}}
\end{center}
\end{figure}

Consider a static timelike photon with momenta $q^\mu=(Q,0,0,0)$ created at $x^\mu=0$ via an $e^+e^-$ annihilation. In the gravity dual description, the subsequent evolution can be pictured as a Kaluza--Klein photon falling down
 into the   $\mbox{AdS}_5$ bulk and reaching the center $z=\infty$  at future infinity $x^+=\infty$. (See Fig.~\ref{fig0} and Refs.~\cite{Hofman:2008ar,Hatta:2008tx}.)
The center of $\mbox{AdS}_5$ is the hyperbolic space $\mbox{H}_3$ parameterized by $W_0^2-W_1^2-W_2^2-W_3^2=1$.
At $x^+=\infty$, or $y^+=0$, the photon hits the point  $W_0=1$, $W_1=W_2=W_3=0$, or in the Poincar\'e 2 coordinates, $y_5=1$ and $y^1=y^2=0$ \cite{Hofman:2008ar}.
Measurements of the final state are done on the boundary ($y^5=0$) of the $\mbox{H}_3$ space which is a sphere parameterized by the solid angle $\Omega=(\theta, \phi$). This is related to $\vec{y}_T$ via the stereographic projection
\begin{align} y^1=\frac{W_1}{W_0+W_3}\approx \frac{\sin \theta \cos \phi}{1+\cos \theta}\,, \quad y^2=\frac{W_2}{W_0+W_3} \approx \frac{\sin \theta \sin \phi}{1+\cos \theta}\,, \label{ste}
 \end{align} and \begin{align}
 (dy^1)^2+(dy^2)^2=\frac{1}{(1+\cos \theta)^2}d\Omega^2\,. \label{ste2} \end{align}

 For the measurement of total energy flowing into a specified region of the solid angle, the relevant operator is \cite{Sveshnikov:1995vi}
 \begin{align} {\mathcal E}(\Omega) \equiv \lim_{r\to \infty}r^2\int_0^{\infty} dx^0\, n_iT^{0i}(x^0,r\vec{n})\,, \end{align}
 where $\vec{n}=(\sin \theta \cos \phi, \sin \theta \sin \phi, \cos \theta)$ is a unit vector and $r^2\equiv x_1^2+x_2^2+x_3^2$.
As observed in \cite{Hofman:2008ar}, this operator takes a particularly simple form in the $y$ coordinates when restricted to  $y^+=0$ (or $x^+=
\infty$). Indeed, from  (\ref{bou}) and (\ref{ste}) it follows that \begin{align} r^2=\frac{1}{2(y^+)^2(1+\cos \theta)^2}\,,\quad dx^0=\frac{1+\cos \theta}{\sqrt{2}} dy^-\,, \end{align}
  and\footnote{Note that the conformal transformation $x^\mu \to y^\mu$ involves a Weyl rescaling of $T_{\mu\nu}$ by a factor
  $2(y^+)^2$ as can be seen from (\ref{bou}). The same applies to the charge operator $j_{\mu}$ below. }  \begin{align} n_iT^{0i}(x)=\frac{4(y^+)^2T_{--}(y)}{(1+\cos \theta)^2}\,, \end{align}
 so that \begin{align} {\mathcal E}(\Omega) =\frac{\sqrt{2}}{(1+\cos \theta)^3}\int dy^- T_{--}(y^+=0,y^-,\vec{y}_T) \equiv \frac{1}{(1+\cos \theta)^3} {\mathcal E}(\vec{y}_T)\,. \label{energy} \end{align}

The factor $1/(1+\cos \theta)^3$ indicates that ${\mathcal E}$ has dimension three under the transformation (\ref{ste}).

 Similarly, if the initial state is charged with respect to a U(1) subgroup of some global symmetry group (such as the ${\mathcal R}$--symmetry), one can measure the total charge flowing in the direction $\vec{n}$ by the operator
 \begin{align} {\mathcal Q}(\Omega) \equiv \lim_{r\to \infty}r^2\int_0^{\infty} dx^0\, n_ij^{i}(x^0,r\vec{n})\,, \end{align}
  which in the $y$ coordinates reads
 \begin{align} {\mathcal Q}(\vec{y}_T) \equiv \int_{-\infty}^{\infty} dy^- j_-(y^+=0,y^-,\vec{y}_T)=(1+\cos \theta)^2
{\mathcal Q}(\Omega)\,. \label{charge}  \end{align}

In the Poincar\'e 2 coordinates, one can imagine that the photon passes through the $\mbox{H}_3$ space at $y^+=0$ and continues to propagate for $y^+>0$.
 Since the operators (\ref{energy}) and (\ref{charge}) are localized at $y^+=0$, they carry  large $p^-$. Consequently, they are sensitive only to the plus component of the original photon momentum
 $q^+=Q/\sqrt{2}$.\footnote{It may seem one is confusing momenta in the $x^-$ and $y^-$ directions. But it turns out that they are the same, see \cite{Hofman:2008ar}.} Therefore, for the purpose of measuring energy and charge at future infinity, one may treat the photon as a shock wave in the $y$--coordinate
 with the following energy momentum tensor
 \begin{align} T_{--}=q^+\delta(y_5-1)\delta^{(2)}(\vec{y}_T)\delta(y^-)\,. \label{shock} \end{align}
  Plugging (\ref{shock}) into the source term of the Einstein equation, one finds the metric \cite{Cornalba:2006xk,Brower:2007qh}
 \begin{align} ds^2=\frac{dy_5^2-2dy^+dy^- +d\vec{y}_T^2+f(y_5,\vec{y}_T)\delta(y^-)(dy^+)^2}{y_5^2}\,, \end{align}
 where
 \begin{align} f(y_5,\vec{y}_T)=\frac{\kappa^2 q^+}{4\pi}y_5 \frac{\left(1+u-\sqrt{u(2+u)}\right)^2}{\sqrt{u(2+u)}}\,. \end{align} In the above, $\kappa^2$ is the five--dimensional gravitational constant
  and $u$ is the chordal distance in $\mbox{H}_3$ measured from the shock wave at $(\vec{y}_T,y^5)=(\vec{0},1)$ \begin{align} u=\frac{(\vec{y}_T)^2+(y_5-1)^2}{2y_5}\,. \end{align}
 The energy momentum at the boundary is given by the well--known procedure
 \begin{align} \langle T_{--}(y^-,\vec{y}_T)\rangle=\lim_{y_5\to 0} \frac{2}{\kappa^2y^4_5} f(y_5,\vec{y}_T)\delta(y^-)= \frac{2q^+}{\pi(1+(\vec{y}_T)^2)^3} \delta(y^-)=\frac{1}{2}T_{00}(y^-,\vec{y}_T)\,, \end{align}
 and therefore,
 \begin{align} \langle{\mathcal E}(\vec{y}_T)\rangle=\sqrt{2}\int_{-\infty}^{\infty} dy^- \langle T_{--}(y^+=0,y^-,\vec{y}_T)\rangle=\frac{2Q}{\pi(1+(\vec{y}_T)^2)^3}\,. \end{align}
  Using (\ref{energy}), one finds the expected result  \begin{align} \langle{\mathcal E}(\Omega)\rangle=\frac{Q}{4\pi}\,, \end{align}
 namely, the distribution of energy is spherical. In \cite{Hofman:2008ar} this result was
 derived in a different way, by evaluating the three--point function $\langle {\mathcal O}^\dagger(q) {\mathcal E}(\Omega){\mathcal O}(q)\rangle$.

Similarly, one can compute the charge distribution. Let us introduce a charge current associated with the shock wave
\begin{align} j^+=\delta(y_5-1)\delta^{(2)}(\vec{y}_T)\delta(y^-)\,, \end{align} and solve the Maxwell equation  in the presence of this current.
\begin{align} \frac{1}{g_{YM}^2}D_{M}F^{M +}=\frac{1}{g_{YM}^2}\frac{1}{\sqrt{-G}}\partial_M\left(\sqrt{-G}F^{M +} \right)=-j^+\,. \end{align}
  Explicitly, \begin{align} \frac{1}{\sqrt{-G}}\partial_M\left(\sqrt{-G}F^{M +} \right)  &=
  -y_5^2\left(y_5^2\partial_i^2 + y_5^2 \partial_{y_5}^2-y_5\partial_{y_5}\right)A_- \nonumber \\ &=-y_5^2 \square_{\mbox{\scriptsize{H}}_3} A_- = -g_{YM}^2j^+\,, \end{align} where $\square_{\mbox{\scriptsize{H}}_3}$ is the scalar Laplacian on $\mbox{H}_3$.
This can be solved using Green's function \cite{Muck:1998rr}
\begin{align} A_-=-\frac{A^+}{y_5^2}&=g_{YM}^2\delta(y^-)\int d^3y' \sqrt{-G'_{\mbox{\scriptsize{H}}_3}}G(y,y')\frac{\delta(y_5'-1)\delta^{(2)}(\vec{y}_T)}{y_5'^2}
\nonumber \\ &=\delta(y^-)\frac{g_{YM}^2}{2\pi} \frac{-1}{(2u)^2} \,_2F_1\left(2,\frac{3}{2},3;-\frac{2}{u}\right)\,. \end{align}
Near the boundary $y_5 \to 0$,
\begin{align} A^+ \approx  \delta(y^-)\frac{g_{YM}^2}{2\pi} \frac{y_5^4}{(1+(\vec{y_T})^2)^2}\,. \end{align}
The boundary expectation value of the current  is computed in the usual way \cite{Gubser:1998bc}
\begin{align} \langle j^+(y^-,\vec{y}_T)\rangle = \frac{\delta S}{\delta A^-} =-\frac{1}{4g^2_{YM}} \left(-4\sqrt{-G}F^{y_5 +}
\right)_{y_5=0} = \delta(y^-)\frac{1}{\pi }\frac{1}{(1+(\vec{y_T}^2))^2}\,. \end{align}
The integrated transverse charge distribution is
\begin{align} \langle {\mathcal Q}(\vec{y}_T)\rangle= \int dy^- \langle j^+\rangle =\frac{1}{\pi }\frac{1}{(1+(\vec{y_T}^2))^2}\,. \end{align}
 Using (\ref{charge}), one finds that the distribution of charge is also spherical
 \begin{align} \langle {\mathcal Q}(\Omega)\rangle= \frac{1}{4\pi}\,. \end{align}

Therefore, although somewhat counterintuitive, it is possible to describe a static photon as a shock wave in this particular coordinate system. This paves a way to establish a close correspondence between the process studied and  the usual shock wave picture of high energy scattering.

\subsection{A shock wave picture of a high energy hadron}
\begin{figure}
\begin{center}
\centerline{\epsfig{file=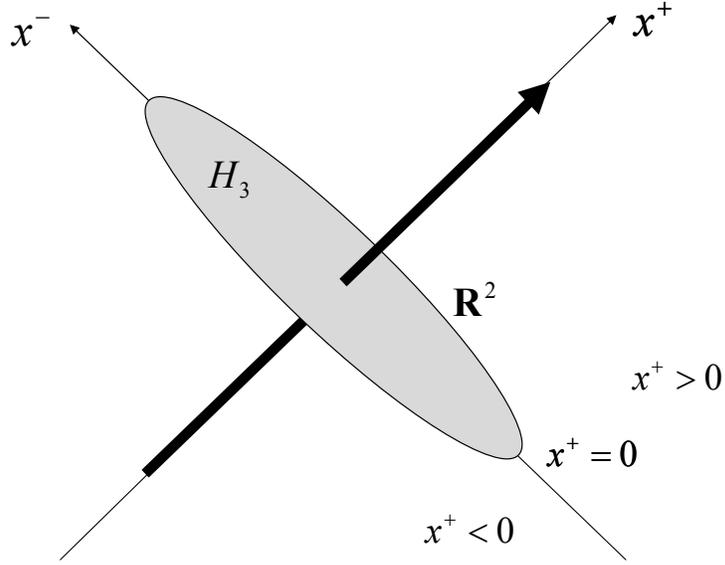,height=7.5cm,width=9.5cm}}
\caption{\sl
 A shock wave picture of a high energy hadron propagating in the $x^+$ direction.
 \label{figure}}
\end{center}
\end{figure}

Consider a state with large momentum $p^+=(p^0+p^3)/\sqrt{2}=\sqrt{2}E$ in the $x^3$ direction created by a gauge invariant operator in ${\mathcal N}=4$ SYM at strong coupling. If this state (`hadron') has a characteristic
 transverse scale $L$, the wavefunction in the $z$--direction $\Phi(z)$ that solves the equation of motion in $\mbox{AdS}_5$ would be localized at around $z\sim L$. At high energy, this portion of the wavefunction can be represented by a shock wave in the Poincar\'e 1 coordinates   \begin{align} T^{++}=\frac{zp^+}{\sqrt{-G}\sqrt{-g_{00}}}\delta(z-L)\delta^{(2)}(\vec{x}_T)\delta(x^-)=z^7p^+\delta(z-L)\delta^{(2)}
 (\vec{x}_T)\delta(x^-)\,, \label{shock2} \end{align} where $zp^+$ is the energy in local inertial coordinates. It is understood that  outcomes of this approximation are to be convoluted with the weight $|\Phi(L)|^2$  in order to obtain physical amplitudes.
   (\ref{shock2}) is identical to (\ref{shock}) after relabeling $y\to x$, $y^5 \to z$, or in other words, going from Poincar\'e 2 to Poincar\'e 1. Moreover, the analog of the condition $y^+=0$ in the $e^+e^-$ case naturally emerges  here due to the Lorentz time dilation. Because the wavefunction of a hadron does not depend on $x^+$, without loss of generality one can decide to work in the $\mbox{H}_3$ space $W_{-1}^2-W_1^2-W_2^2-W_4^2=1$ located at $x^+=0$. The boundary of this $\mbox{H}_3$ space at  $z=0$ can be identified with the transverse (or the impact parameter) plane $\mathbb{R}^2$, $\vec{x}_T=(x^1,x^2)$ in the physical Minkowski space, see Fig.~(\ref{figure}). It is now obvious that the calculation of energy and charge distributions [The former was initially done in \cite{Gubser:2008pc} which we reproduce below.] is mathematically identical to the one given in the previous subsection.\footnote{The only modifications are the replacement $1\to L$ and the factors of $\sqrt{2}$ in the energy--momentum tensor arising from the different initial four--momenta, namely, $q^\mu=(Q,0,0,0)$ vs. $p^\mu=(E,0,0,E)$. }

 (\ref{shock2}) modifies the metric as
 \begin{align} ds^2=\frac{dz^2-2dx^+dx^- +d\vec{x}_T^2+f(z,\vec{x}_T)\delta(x^-)(dx^+)^2}{z^2}\,, \end{align}
 where
 \begin{align} f(z,\vec{x}_T)=\frac{\kappa^2 p^+}{4\pi}zL \frac{\left(1+u-\sqrt{u(2+u)}\right)^2}{\sqrt{u(2+u)}}\,, \end{align}
   and \begin{align} u=\frac{(\vec{x}_T)^2+(z-L)^2}{2zL}\,. \end{align} At high energy the dominant component of the energy momentum tensor on the boundary is $T^{++}=T_{--}=2T^{00}$, and one is typically interested in the transverse energy distribution \begin{align}
   {\mathcal E}(\vec{x}_T) =\frac{1}{\sqrt{2}}\int_{-\infty}^{\infty} dx^-  T_{--}(x^+=0,x^-,\vec{x}_T)\,. \end{align} This is identical in structure to the corresponding operator (\ref{energy}) in the timelike problem.
 Its expectation value is given by
 \begin{align}  \langle T_{--}(x^-,\vec{x}_T)\rangle=\lim_{z\to 0} \frac{2}{\kappa^2z^4} f(z,\vec{x}_T)\delta(x^-)
  =\frac{2p^+L^4}{\pi(L^2+(\vec{x}_T)^2)^3} \delta(x^-)\,,  \end{align}
  and therefore,
 \begin{align} \langle {\mathcal E}(\vec{x}_T)\rangle =\frac{1}{\sqrt{2}}\int_{-\infty}^{\infty} dx^- \langle T_{--}(x^+=0,x^-,\vec{x}_T)\rangle =\frac{2EL^4}{\pi(L^2+(\vec{x}_T)^2)^3}\,.  \label{tran} \end{align}
 The total energy is
 \begin{align} \int d^2 \vec{x}_T \, \langle{\mathcal E}(\vec{x}_T)\rangle=E\,, \end{align}
 as it should.

 Incidentally, consider also the Fourier transform
 \begin{align} \langle{\mathcal E}(\vec{q}_T)\rangle=\int d^2\vec{x}_T \, e^{i\vec{q}_T\cdot \vec{x}_T} \langle{\mathcal E}(\vec{x}_T)\rangle=\frac{EL^2}{2}q_T^2K_2(q_TL)\,. \label{had} \end{align} This is the `gravitational form factor' of our pointlike probe localized at $z=L$ \cite{Abidin:2008sb,Brodsky:2008pf}.  The form factor of realistic hadrons can be obtained by convoluting (\ref{had}) with a suitable
 wavefunction. Then the exponential behavior $\sim e^{-q_TL}$ at large $q_T$ will be replaced by the usual power--law falloff.


Similarly, starting from the charge current associated with the shock wave
\begin{align} j^+=\frac{1}{\sqrt{-G}}\delta(z-L)\delta^{(2)}(\vec{x}_T)\delta(x^-)=z^5\delta(z-L)\delta^{(2)}(\vec{x}_T)\delta(x^-)\,, \end{align}
 one finds
the transverse charge distribution
\begin{align} \langle{\mathcal Q}(\vec{x}_T)\rangle= \int dx^-\langle j^+\rangle =\frac{1}{\pi }\frac{L^2}{(L^2+(\vec{x_T}^2))^2}\,,
 \end{align} which is correctly normalized \begin{align} \int d^2\vec{x}_T \langle{\mathcal Q}(\vec{x}_T)\rangle=1\,.
 \end{align}
 Its Fourier transform is
 \begin{align}\langle {\mathcal Q}(\vec{q}_T)\rangle=\int d^2\vec{x}_T \, e^{i\vec{q}_T\cdot \vec{x}_T} \langle {\mathcal Q}(\vec{x}_T)\rangle=
 q_TLK_1(q_TL)\,.
  \end{align}
 This is  the `electromagnetic form factor' of our probe. See  \cite{Brodsky:2006uqa,Grigoryan:2007vg} for more discussions.

To sum up, we have shown that there is an exact mapping between seemingly different physical problems at strong coupling: the final state in $e^+e^-$ annihilation and the transverse profile of a colorless state at high energy. Since the relevant operators are identical up to the relabeling $x\leftrightarrow y$ of their arguments, the correspondence holds also for the correlation functions  of these operators, including stringly corrections (finite--$\lambda$ effects). We will discuss this issue later, but before doing so let us turn to the weak coupling case and see if there is a similar correspondence between the two processes.

\section{Interjet soft gluon multiplicity at weak coupling}
\setcounter{equation}{0}
As explained in Introduction, at weak coupling the only place one may hope to look for Pomeron--like dynamics in $e^+e^-$ annihilation is the interjet (away--from--jet) region where only the soft logarithms are relevant. In this section we shall investigate only the multiplicity and distribution of soft gluons, though our discussion is general and likely to be applicable to other non--global observables.  It is convenient to work in the large $N_c$ approximation\footnote{This is of course implicit in the AdS/CFT--based calculations in the previous section.} and use the color dipole approach \cite{Gustafson:1987rq,Mueller:1993rr}.
In this picture, the emission of a soft gluon with momentum $k$ from the primary quark and antiquark with momenta $p_a$ and $p_b$, respectively,  can be viewed as a splitting of one dipole into two child dipoles. The differential probability for this splitting is given by
 \begin{align} dP=\bar{\alpha}_s \omega d\omega \frac{d\Omega_k}{4\pi}\frac{p_a\cdot p_b}{(p_a \cdot k)(k\cdot p_b)} \approx \bar{\alpha}_s \frac{d\omega}{\omega} \frac{d\Omega_k}{4\pi}\frac{1-\cos \theta_{ab}}{(1-\cos \theta_{ak})(1-\cos \theta_{bk})}\,, \label{spl} \end{align} where $\bar{\alpha}_s=N_c\alpha_s/\pi$, $\omega=k^0$ is the energy of the soft gluon and $\theta_{ak}$ is the angle between the quark and  gluon directions, etc.

 As the invariant mass $Q$ is increased, the initial $q\bar{q}$ pair undergoes many splittings and produces a lot of dipoles (gluons) with a small fraction of energy $\omega/Q\ll 1$.
 Defining $n(\theta_{ab},\theta_{cd},Y)$  as the number density of dipoles with opening angle $\theta_{cd}$ inside the parent dipole with opening angle $\theta_{ab}$ within the rapidity interval $Y=\ln Q/\omega$, Ref.~\cite{Marchesini:2003nh} derived the following evolution equation\footnote{By slight abuse of notation, $\theta_{ab}$ denotes both the coordinates  $\Omega_{a(b)}=(\theta_{a(b)},\phi_{a(b)})$ of the dipole legs on a sphere as well as the magnitude of the angle between. Similarly, $x_{ab}$ etc. in (\ref{bf}) denotes both the two dimensional vectors $\vec{x}_{a(b)}$ as well as their magnitude $|\vec{x}_a-\vec{x}_b|$. }
 \begin{align} \partial_Y n(\theta_{ab},\theta_{cd}, Y)=\bar{\alpha}_s\int \frac{d^2\Omega_k}{4\pi} \frac{1-\cos \theta_{ab}}{(1-\cos \theta_{ak})(1-\cos \theta_{bk})} \nonumber \\
 \times \bigl(n(\theta_{ak},\theta_{cd},Y)+n(\theta_{bk},\theta_{cd},Y)-n(\theta_{ab},\theta_{cd},Y)\bigr)\,. \label{for} \end{align}

  Turning now to the spacelike problem, one can contrast (\ref{for})  with the evolution equation for the dipole density $n(x_{ab},x_{cd})$ in the transverse plane defined as the total number of dipoles with size $x_{cd}$ and energy larger than $\omega$ inside the parent dipole with size $x_{ab}$ and energy $E$ \begin{align} &\partial_Y n(x_{ab},x_{cd}, Y)=\bar{\alpha}_s\int \frac{d^2\vec{x}_k}{2\pi} \frac{(\vec{x}_{ab})^2}{(\vec{x}_{ak})^2(\vec{x}_{bk})^2}  \nonumber \\
& \qquad \quad \times \bigl(n(x_{ak},x_{cd},Y)+n(x_{bk},x_{cd},Y)-n(x_{ab},x_{cd},Y)\bigr)\,, \label{bf} \end{align} where $Y=\ln E/\omega$.

 In \cite{Marchesini:2003nh}, it was argued that if one considers $e^+e^-$ annihilation in a highly boosted frame such that $\theta_{ab}\ll 1$,  one can approximate $1-\cos \theta \approx \theta^2/2$ and $d^2\Omega
\approx d^2\vec{\theta}$. Then the two equations (\ref{for}) and (\ref{bf}) become formally identical. This means, in particular, that the leading energy dependence is that of the BFKL Pomeron \begin{align} n(\theta_{ab},\theta_{cd}, Y) \sim n(x_{ab},x_{cd}, Y)  \sim e^{4\bar{\alpha}_s\ln2 \,Y}\,. \end{align}
 A more detailed analysis of (\ref{for}) without assuming $\theta_{ab}$ to be small has been carried out in  \cite{Marchesini:2004ne}. There the dependence on $\theta_{cd}$ was suppressed, so the result  is inclusive with respect to the size and location of the child dipoles.

\begin{figure}
\begin{center}
\centerline{\epsfig{file=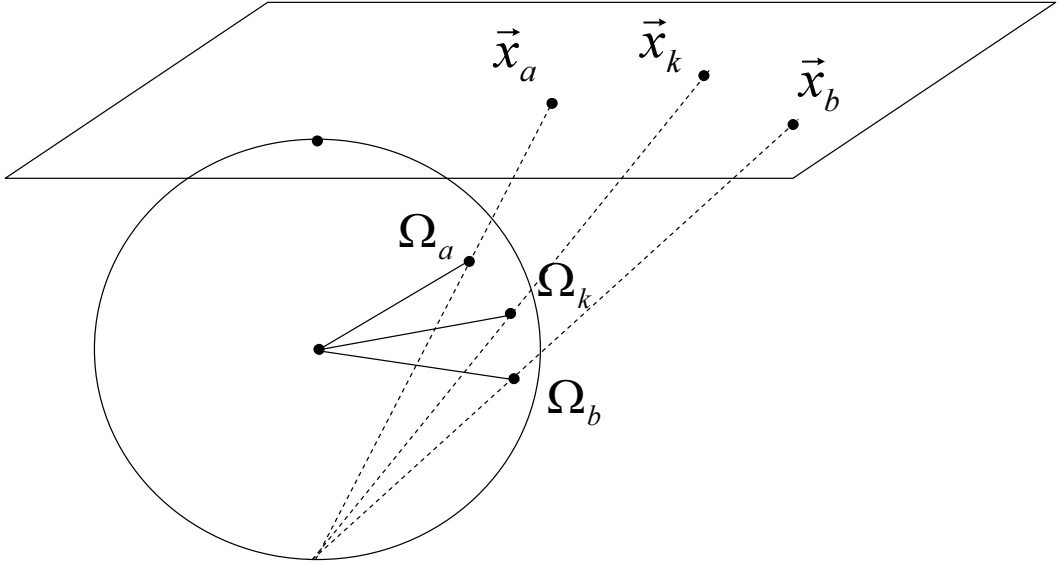,height=7.5cm,width=13.cm}}
\caption{\sl
 Stereographic map between a sphere of unit diameter  and the transverse plane.
 \label{fig1}}
\end{center}
\end{figure}

 Here we wish to point out that the two equations with the full $\theta_{cd}$ and $x_{cd}$ dependence, respectively, can be \emph{exactly} mapped to each other via the stereographic projection
 (\ref{ste}) and (\ref{ste2}), see, Fig.~\ref{fig1}. Indeed, (\ref{ste}) implies (renaming $y\to x$ and omitting the subscript $T$) \begin{align}
 \cos \theta=\frac{1-|\vec{x}|^2}{1+|\vec{x}|^2}, \quad \sin \theta=\frac{2|\vec{x}|}{1+|\vec{x}|^2}, \quad \cos \phi=\frac{x^1}{|\vec{x}|}, \quad
 \sin \phi=\frac{x^2}{|\vec{x}|}\,, \end{align}
  from which it follows that \begin{align} 1-\cos \theta_{ab}=\frac{2(\vec{x}_{ab})^2}{(1+(\vec{x}_a)^2)(1+(\vec{x}_b)^2)}\,, \label{transf} \end{align}
   and \begin{align}
   \frac{d^2\Omega_k}{4\pi} \frac{1-\cos \theta_{ab}}{(1-\cos \theta_{ak})(1-\cos \theta_{bk})}
   =\frac{d^2\vec{x}_k}{2\pi} \frac{(\vec{x}_{ab})^2}{(\vec{x}_{ak})^2(\vec{x}_{bk})^2}\,.  \label{tan}\end{align}
 Moreover, the initial conditions for the two equations are identical, namely, at $Y=0$ the parent and the child dipole are on top of one another; $\theta_{ab}=\theta_{cd}$ and $x_{ab}=x_{cd}$.  Since the exact solution to $n(x_{ab},x_{cd}, Y)$ is known  \cite{Mueller:1994gb,Navelet:1997tx}, all one needs to do is to make the coordinate transformation (\ref{transf}) to obtain the exact solution $n(\theta_{ab},\theta_{cd}, Y)$. This task becomes very straightforward once one notices that, thanks to conformal symmetry of the BFKL equation, the solution depends only on the anharmonic ratio which takes a  simple form in each coordinate system \begin{align} |\rho|^2=\frac{x_{ab}^2x_{cd}^2}{x_{ac}^2x_{bd}^2}=\frac{(1-\cos \theta_{ab})(1-\cos\theta_{cd})}{(1-\cos \theta_{ac})(1-\cos\theta_{bd})}\,. \end{align}

Instead of writing down the exact solution for generic values of $|\rho|$ and $Y$, which is possible but complicated, let us focus on the most interesting case of back--to--back jets $\theta_{ab}=\pi$ and count the number of dipoles with $\theta_{cd}\ll 1$ in the interjet region $\theta_{ac}\approx \pi-\theta_{bd} \sim {\mathcal O}(1)$. Then \begin{align}  |\rho|\approx \frac{\theta_{cd}}{\sin \theta_{ac}} \ll 1\,.  \label{small} \end{align}
 After the stereographic projection, this is equivalent to the number of small dipoles $x_{cd} \ll x_{ab}$ near the center of the parent $\frac{\vec{x}_a+\vec{x}_b}{2}$.
The asymptotic behavior at large $Y$ is\footnote{Our normalization of $n(x_{cd})$ differs from the more common one in the literature by a factor $1/x_{cd}^2$. The current normalization directly gives the total number of dipoles upon integration with the measure $\int d^2\vec{x}_{cd}$ rather than $\int d^2 \vec{x}_{cd}/x_{cd}^2$.}
\begin{align} x_{cd}^4n(x_{ab},x_{cd}, Y) \sim \theta_{cd}^4n(\theta_{ab},\theta_{cd}, Y) \sim \frac{|\rho|}{(D\bar{\alpha}_sY)^{3/2}}\ln \left(\frac{16}{|\rho|}\right)\  e^{4\ln 2 \bar{\alpha}_s Y} e^{-\frac{2\ln^2 (|\rho|/16)}{D\bar{\alpha}_sY}}\,,  \label{inter} \end{align} with $D=28 \zeta(3)$.
 The most salient feature of (\ref{inter})
is the nontrivial angular distribution of dipoles/gluons in the interjet region \begin{align} n(\theta_{ab},\theta_{cd},Y) \propto \frac{1}{\sin \theta_{ac}}\,, \label{intu}\end{align} which cannot be obtained in the small angle approximation $\theta_{ab}\ll 1$, $\sin \theta_{ac} \approx \theta_{ac}$.
 (\ref{intu}) is perfectly in line with one's physical intuition that the multiplicity distribution has peaks in the directions of the back--to--back jets $\theta_{ac}\approx 0$ and $\theta_{ac}\approx \pi$. On the other hand, to obtain the total multiplicity within jets one has to analyze (\ref{for}) taking into account  double logarithmic contributions \cite{Dokshitzer:2008ia}.

\section{Correlation functions}
\setcounter{equation}{0}
So far we have been focusing on the timelike--spacelike correspondence of energy, charge and number distributions.
 In this last section we briefly discuss how the correspondence works at the level of higher--point functions of these operators.

At strong coupling,
Ref.~\cite{Hofman:2008ar} has given a prescription to compute  correlation functions of the energy operator (\ref{energy}) \begin{align}
\langle {\mathcal E}(\Omega){\mathcal E}(\Omega')\cdots \rangle\,, \end{align} beyond the supergravity approximation.
Up to  the coordinate transformation (\ref{ste}) and the rescaling (\ref{energy}), they give correlation functions of ${\mathcal E}(\vec{y})$'s on the $y$--plane
\begin{align}  \langle {\mathcal E}(\vec{y}_a){\mathcal E}(\vec{y}_b)\cdots \rangle\,. \label{tty}\end{align}
From our point of view the only difference between the timelike and spacelike problems is just the relabeling $ x \leftrightarrow y$. Thus one can immediately get from (\ref{tty})  correlation functions of the energy operator (\ref{tran}) in the transverse plane.
\begin{align} \langle {\mathcal E}(\vec{x}_a){\mathcal E}(\vec{x}_b)\cdots \rangle=\langle {\mathcal E}(\vec{y}_a){\mathcal E}(\vec{y}_b)\cdots \rangle\arrowvert_{\vec{y}\to \vec{x}}\,, \end{align} measured in the presence of a high energy
 hadron.  In particular, the complete absence of correlation in the $\lambda \to \infty$ limit \cite{Hofman:2008ar}
   \begin{align} \langle \prod_i{\mathcal E}(\vec{x}_i) \rangle = \prod_i \langle {\mathcal E}(\vec{x}_i) \rangle\,, \end{align}
   should still be  valid here. Another interesting result is the behavior in small angle limit
  \begin{align} \langle {\mathcal E}(\Omega_a){\mathcal E}(\Omega_b) \rangle \sim \frac{1}{|\theta_{ab}|^{2+2\gamma(3)}}\,, \quad (\theta_{ab}\to 0) \label{in}  \end{align} where $\gamma(j)$ is the (spacelike) anomalous dimension.
 In the hadron problem, this should translate into
   \begin{align} \langle {\mathcal E}(\vec{x}_a){\mathcal E}(\vec{x}_b) \rangle \sim \frac{1}{|\vec{x}_a-\vec{x}_b|^{2+2\gamma(3)}}\,. \label{ma} \end{align} Since (\ref{in}) is an exact field theoretic result,  it is valid in ${\mathcal N}=4$ SYM for any value of the coupling. At strong coupling, the anomalous dimension is large and negative, $\gamma(3)\approx -\lambda^{1/4}/\sqrt{2}$, so the correlation vanishes as $\Omega_b\to \Omega_a$ or $\vec{x}_b \to \vec{x}_a$. In view of the result in Section~3, one can expect (\ref{in}) and (\ref{ma}) to be valid also in QCD to the extent that the running of the coupling can be neglected (as in the leading order BFKL approximation).  To see this is indeed the case, return to (\ref{for}) and express the solution in the inverse Mellin representation keeping  $\theta_{cd} \ll 1$
  \begin{align} n(\theta_{ab},\theta_{cd},Y)\sim \frac{1}{\theta^2_{cd}}\int \frac{d\gamma}{2\pi i}e^{(j(\gamma)-1)Y}\left(\frac{1}{\theta_{cd}}
  \right)^{2\gamma}= \frac{1}{\theta^2_{cd}}\int \frac{dj}{2\pi i}c(j)\left(\frac{1}{x}\right)^{j-1} \left(\frac{1}{\theta_{cd}}
  \right)^{2\gamma(j)}\,,  \end{align}
  where
   \begin{align} j(\gamma)-1=\bar{\alpha}_s \left(2\psi(1)-\psi(\gamma)-\psi(1-\gamma)\right)\,, \label{sol} \end{align}
  $c(j)$ is the Jacobian and $x=e^{-Y}$ is the usual small--$x$ variable (not to be confused with  transverse coordinates such as $x_a$.). In order to relate (\ref{sol}) to the energy two--point function with opening angle $\theta_{cd}$,
   one multiplies by $x$ and integrates over $x$\footnote{Note that $n(x)$ is the total number of dipoles with energy fraction larger than $x$. Therefore, one $x$--integration is already implied in its definition.}
  \begin{align} \int_0^1 dx \,x n(\theta_{ab},\theta_{cd},x) \sim \frac{1}{\theta^2_{cd}}\int dj \int_0^1 dx\,x^{2-j}\left(\frac{1}{\theta_{cd}}
  \right)^{2\gamma(j)}  \nonumber \\
  \sim \frac{1}{\theta_{cd}^2}\int dj \frac{1}{j-3} \left(\frac{1}{\theta_{cd}}
  \right)^{2\gamma(j)}  \sim \frac{1}{\theta_{cd}^{2+2\gamma(3)}}\,. \label{sug} \end{align}
     This has the same singularity structure as (\ref{in}). In the BFKL approximation, $\gamma(3)$ is obtained by solving (\ref{sol}) with $j=3$ in the branch $0< \gamma< 1/2$.  (\ref{sug}) suggests that the opening angle $\theta_{cd}$ of child dipoles in the energy--weighted distribution can be reinterpreted as the separation angle in the energy--energy correlation function at least when $\theta_{cd}$ is small.
 Similarly, in the hadron problem at weak coupling one finds, as $x_{cd} \to 0$,
  \begin{align}
 \int_0^1 dx \,xn(x_{ab},x_{cd},x)  \sim \frac{1}{x_{cd}^{2+2\gamma(3)}}\,,    \end{align}
 in agreement with (\ref{ma}).

\section{Conclusions}
We have pointed out a novel correspondence between the interjet region in $e^+e^-$ annihilation and  the transverse plane of a high energy hadron both at weak and strong coupling.  At strong coupling, due to the absence of jets the mapping extends to the entire region $S^2 \leftrightarrow {\mathbb R}^2$ where particles are distributed. The fact that the same stereographic projection works both at weak and strong coupling suggests that the timelike--spacelike duality in the soft sector has a deep geometric origin which persists beyond the perturbative framework. We have demonstrated the usefulness of this mapping for the single dipole distribution.
 Recently there has been some progress in the analytical and numerical study of multiple dipole correlation functions \cite{Hatta:2007fg,Xiao:2007te,Avsar:2008ph}. The corresponding correlation of dipoles in $e^+e^-$ annihilation will be studied elsewhere.

\section*{Acknowledgments}
I thank Emil Avsar, Toshihiro Matsuo  for discussions and Al Mueller for comments.
This work is supported, in part, by Special Coordination Funds for Promoting Science and Technology of the Ministry of Education, Culture, Sports, Science and Technology, the Japanese Government.

\end{document}